# Automated identification and characterization of parcels (AICP) with OpenStreetMap and Points of Interest

Ying Long, Beijing Institute of City Planning, China, longying1980@gmail.com
Xingjian Liu, The University of Hong Kong, China, xliu6@hku.hk (the corresponding author)

**Abstract**: Against the paucity of urban parcels in China, this paper proposes a method to automatically identify and characterize parcels (AICP) with OpenStreetMap (OSM) and Points of Interest (POI) data. Parcels are the basic spatial units for fine-scale urban modeling, urban studies, as well as spatial planning. Conventional ways of identification and characterization of parcels rely on remote sensing and field surveys, which are labor intensive and resource-consuming. Poorly developed digital infrastructure, limited resources, and institutional barriers have all hampered the gathering and application of parcel data in developing countries. Against this backdrop, we employ OSM road networks to identify parcel geometries and POI data to infer parcel characteristics. A vector-based CA model is adopted to select urban parcels. The method is applied to the entire state of China and identifies 82,645 urban parcels in 297 cities. Notwithstanding all the caveats of open and/or crowd-sourced data, our approach could produce reasonably good approximation of parcels identified from conventional methods, thus having the potential to become a useful supplement.

**Introduction**

Land parcel data are one of the cornerstones of contemporary urban planning (Cheng et al. 2006)[1]. Parcels are the basic spatial units of urban models, as for example the latest urban simulation models are oftentimes vector-based and capture parcel-level dynamics (Stevens and Dragicevic 2007; Pinto 2012). More importantly, normative planning and policies are performed on parcels, ranging from the formulation of comprehensive plans, to strategic plan implementation, and to policy evaluation (Frank et al. 2006; Jabareen 2006; Alberti et al. 2007).

Whereas parcel data for the developed world are generated by robust digital infrastructure and supplemented by open data initiatives (e.g., OpenStreetMap), researchers still lament the difficulty of attaining parcel data for developing countries. For example, the best available parcel map for China's capital Beijing – supposedly one of the most technologically advanced and rapidly developing cities in the erstwhile Third World – was dated in 2010 (Beijing Institute of City Planning 2010). In addition, collecting parcel data in medium- and small- sized cities in China is constrained by poorly developed digital infrastructures. That goes without saying that complete parcel-level features (e.g., land use type, urban functions, and development density) do not exist on many occasions. In addition to hard infrastructures, soft institutions have also created barriers for Chinese urban planners' access to parcel maps. For instance, our interviews with 57 planning professionals[2] suggest that the access to existing parcel maps held by local planning bureaus/institutes is extremely restrained, as parcel maps are tagged as confidential within the current Chinese planning institutions. In summary, parcel data for the developing world are oftentimes outdated, limited in geographical scopes, and do not contain much thematic information other than basic parcel geometry. As parcel data are at the central stage of urban planning (Cheng et al. 2006), the lack of parcel data would constrain our ability to trace urban changes at high spatial resolution, hinder the formulation and implementation of detailed urban plans, and restrain the possibility of adopting contemporary parcel-based urban management.

---

[1] Parcels in China correspond to "blocks" in the US context. In this paper, we use the term "parcel" to be consistent with other literature on Chinese cities.

Built on manual interpretation of remote sensing images and field surveys, conventional ways of generating parcel data are time consuming, expensive, and labor-intensive (Erickson et al. 2013). Thus many developing countries do not have the necessary capital and resources to produce parcel data in the conventional fashion. Overcoming such data paucity seems to be of high priority for urban planning in developing countries.

Against this backdrop, we propose a method for automatic identification and characterization of parcels (AICP), based on ubiquitous OpenStreetMap (OSM) and Points-of-Interest (POI) data. The proposed method could (1) provide quick and robust delineation of land parcels; and (2) generate a variety of parcel level attributes, allowing for the examination of urban functions, development density and mixed land uses. We illustrate the usefulness of our method with 297 cities in China. In addition, the framework could work well with conventional data sources (e.g., survey data), when the latter are available. The next section reviews the progress in obtaining parcel-level geometry and features, followed by an elaboration of methods and the case study. We conclude with a discussion of the strength, limitations, as well as future applications of our method.

---

[2] We have interviewed 57 planning professionals in China (23 of these 57 professionals are affiliated with research institutes and universities; 21 with foreign planning/architect firms such as AEOCOM and Atkins; and 13 with domestic planning institutes and firms). Professionals often rely on manual digitalization of land use maps (often in raster formats as vector data files would not be released). This process is extremely time consuming and often produces land parcel maps of less desirable quality.

## Identification and characterization of parcels

Parcel boundaries and their features are conventionally identified through manual interpretation of topographic maps, building maps, field surveys, and high-resolution remote sensing images (Cheng et al. 2006). Such manual operations suffer a series of setbacks. Firstly, manual operations are resource intensive and time consuming. For example, it would take an experienced operator 3-5 hours to identify and infer land use for 35-50 urban parcels covering the area of one square kilometers. Secondly, manual operations often result in inconsistent dataset, as data quality largely depends on the experiences and technical proficiency of individual practitioners. Thirdly, conventional methods are less suitable for routine updates and longitudinal comparison. This becomes more problematic in the face of volatile urban development (e.g., gentrification and urban sprawl) in many developing countries. Still, even though manual operations could identify parcel geometries, the generated parcels usually lack parcel level information, such as density and land use mix. As a case in point, data about parcel density for Beijing, China are limited to the area within the sixth ring road (approximately 13.8% of the whole Beijing Metropolitan Area).

Recent attempts have been made to automatically identify parcel geometries. For example, Yuan et al. (2012) proposed a raster-based approach for parcel delineation based taxi trajectories and POIs. However, Yuan et al. (2012) omitted road space in the delineation of parcels, and the raster-based nature of the method generates heavy computational burden, severely limiting the method's applicability to large geographical areas. Meanwhile, Aliagaet al. (2008) presented an algorithm of interactively synthesizing parcel layouts for to-be-developed areas based on the structure of real-world urban areas. This study is limited by the fact that it does not account for parcel characteristics, and performs parcel subdivision within predefined blocks, instead of identifying blocks from the data.

In light of this situation, OSM has been proposed as a promising candidate for quick and robust delineation of parcels and other urban features (Haklay and Weber 2008; Over et al. 2010; Ramm 2010). As one of the most successful volunteered GIS projects, OSM provides street network data for a wide array of cities (Goodchild 2007; Sui 2008). Jokar Arsanjani et al. (2013a) predicted that the data coverage and quality of OSM will continue to be improved in the coming years. More specifically, the quality of OSM data in well-mapped and oftentimes large cities is on par with that of topographic maps (Girres and Touya 2010; Haklay 2010; Over et al. 2010). The growth of OSM in developing countries has been encouraging, as the volume of OSM data in China has experienced a nine-fold increase during 2007-2013 (Figure 1).

Figure 1 Increasing data volume in the OSM-China data (accessed on Oct 5, 2013; Unit: data points).

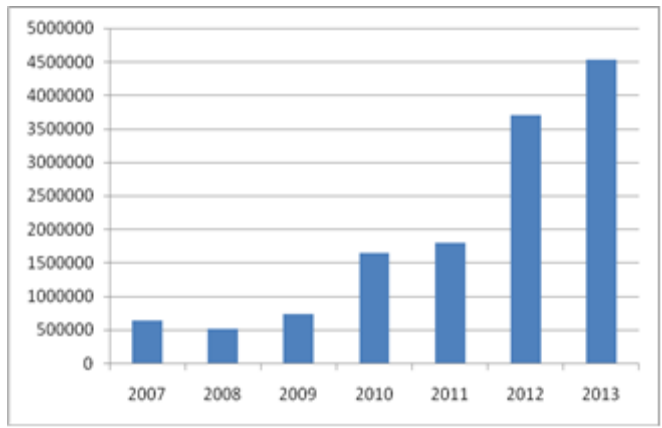

Several preliminary studies suggest that OSM road networks are useful in identifying urban structures. For example, Hagenauer and Helbich (2012) extracted urban built-up areas from OSM, and Jiang and Liu (2012) identified natural grouping of city blocks based on OSM data. Existing analyses using OSM focus more on deriving universal laws and social physics (Jiang and Liu 2012) rather than producing data products for urban planning and studies. In a similar vein, Jokar Arsanjani et al. (2013b) identified land-use patterns for central Vienna, Austria (roughly 32 km$^2$) using OSM. Whereas Jokar Arsanjani et al. (2013b) introduced a volunteered geographic information based approach to generate land-use patterns, their approach focuses on developed countries with high-accuracy OSM data, uses pre-defined boundaries for parcel subdivision, and could be extended to generate additional parcel features.

As a corollary, it is understandable that researchers show hesitations in applying OSM-China data, as the data quality is not always clear and more meta-data may be needed. In light of this situation, attempts have been made to assess the quality of OSM data in China (Zheng and Zheng 2014). Preliminary results suggest that *inter alia* (1) OSM data are more dense and applicable in major cities, although the overall data coverage may be poor; (2) data completeness remains the major issue (Goodchild 2008); (3) OSM data in China have been continuously improved. In line with observations, OSM data are gradually used to understand Chinese cities (see for example, Liu et al. 2012; Leitte et al. 2012; Zhang et al. 2013).

In addition to parcel geometries, planning practices also require parcel features such as urban functions and development density. There is a rich literature on inferring land use from remote sensing images (Kressler et al. 2001; Herold et al. 2002). However, as discussed previously, remote sensing images are not suitable for large scale parcel-level analysis, due to *inter alia* data availability and computational burden. Although some automatic or semi-auto techniques have been developed to address urban land-use classification (Herold et al. 2002; Pacifici et al. 2009), it is still difficult to identify certain land use types such as high-density residential areas and commercial areas from remote sensing images. More importantly, remote-sensing based methods often treat parcels as having homogenous land use types, thus not allowing for mixed land use. More recently, researchers have inferred human use of urban space with human mobility data, such as smart card records (Long et al. 2013), mobile phone data (Soto and Frias-Martinez 2011; Toole et al. 2012), and taxi trajectories (Liu et al. 2012; Yuan et al. 2012). Nevertheless, human mobility data are often proprietary and involve privacy issues (Beresford and Stajano 2003). Such limited data access greatly undermines the wide applicability of human-mobility based methods.

To this end, we propose Points of Interest (POI) data as an alternative data source for characterizing parcels. The strength of POI data includes (1) containing sub-parcel level business information, which could serve as proxies for land use and urban functions; (2) being available from online mapping and cataloguing service providers; (3) having a nearly global coverage; and (4) having high spatial (e.g., geo-coded business locations) and temporal (e.g., routinely updated by service providers) resolutions. In addition, while some POI data are proprietary and require small access fees, many are freely available from online business catalogues. With all these advantages, POI data seem to have great potential in characterizing parcel features.

Therefore, we propose a preliminary automatic process to improve the identification and characterization of fine-scale urban land parcels. OSM data are used to identify and delineate parcel geometries, while POIs are gathered to infer land use intensity, function, and mixing at the parcel-level. We emphasize that our empirical framework is (1) fully automatic and extensible, allowing for the incorporation of various data sources (e.g., taxi trajectories, mobile phone data, as well as conventional survey data); (2) produce not only parcel geometry and land use types but also useful parcel-level information such as land use mix; (3) is applicable to large geographic areas, while most previous studies are limited to small areas; and (4) enables routine updates and free distribution of urban parcel data for China. While enjoying the ubiquitous availability and other blessings of OSM data, the application of our method needs to be aware of the caveats of crowd-sourced and/or open data (Elwood et al. 2012; Neis et al. 2012; Sui et al. 2013).

## Data and methods

**Data**

*Administrative boundaries of Chinese cities*

Our analysis covers a total of 654 cities in China (Figure 2)[3], ranging across five administrative levels: municipalities directly under the Central Government (MD, 4 cities), sub-provincial cities (SPC, 15), other provincial capital cities (OPCC, 16), prefecture-level cities (PLC, 251), and county-level cities (CLC, 368) (Ministry of Housing and Urban Development, MOHURD, 2013; see Ma, 2005 for a more detailed discussion on the Chinese administrative system). As a city *proper* in China contains both rural and urban land uses, we narrow our analytical scope onto legally defined urban land within city *propers* and use administrative boundaries of urban lands to carve out OSM and POI data layers. In addition to administrative boundaries, we also gather information about total build-up area of individual cities in 2012 (MOHURD, 2013), which will be used in the urban parcel identification process.

Figure 2 Administrative boundaries of Chinese cities

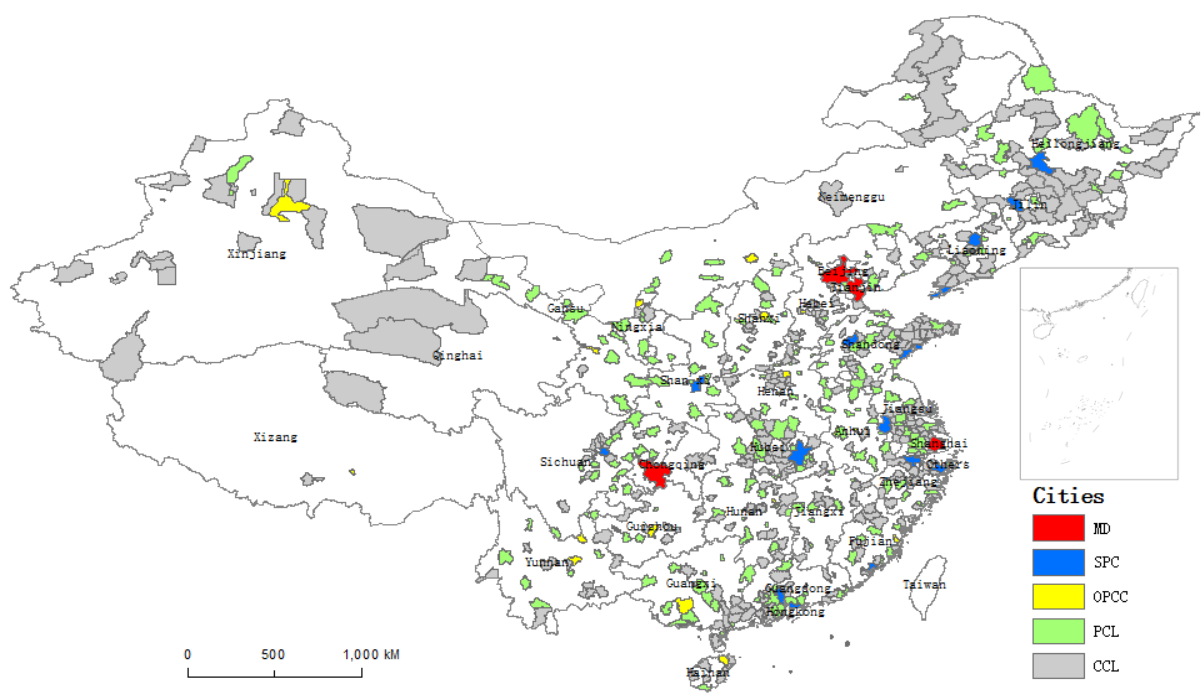

---

[3] Sansha in Hainan, Beitun in Xinjiang, and Taiwan were not included in our analysis due to the availability of OSM and POI data.

*OSM in China*

OSM road networks for China were downloaded on October 5, 2013. We also amass the ordnance survey data (ORDNANCE) of China at the end of 2011 with detailed road networks to verify results produced by OSM data. The OSM dataset contains 481,647 road segments (8.0% of that of the ordnance survey map) of 825,382 kilometers (31.5% of the ordnance survey map). Furthermore, road networks in OSM and the ordnance survey map are overlaid for a visual inspection of data quality (Figure 3). The preliminary completeness check is consistent with previous findings on OSM data quality in China (Zheng and Zheng, 2014). In spite of capturing a portion of the ordnance survey data, OSM data cover most urban areas in China, especially large cities (Figure 3), and are potentially useful for identifying urban land parcels. The implication of OSM data quality will be further elaborated in subsequent sections.

Figure 3 A comparison of roads in the OSM and ordnance map[4]

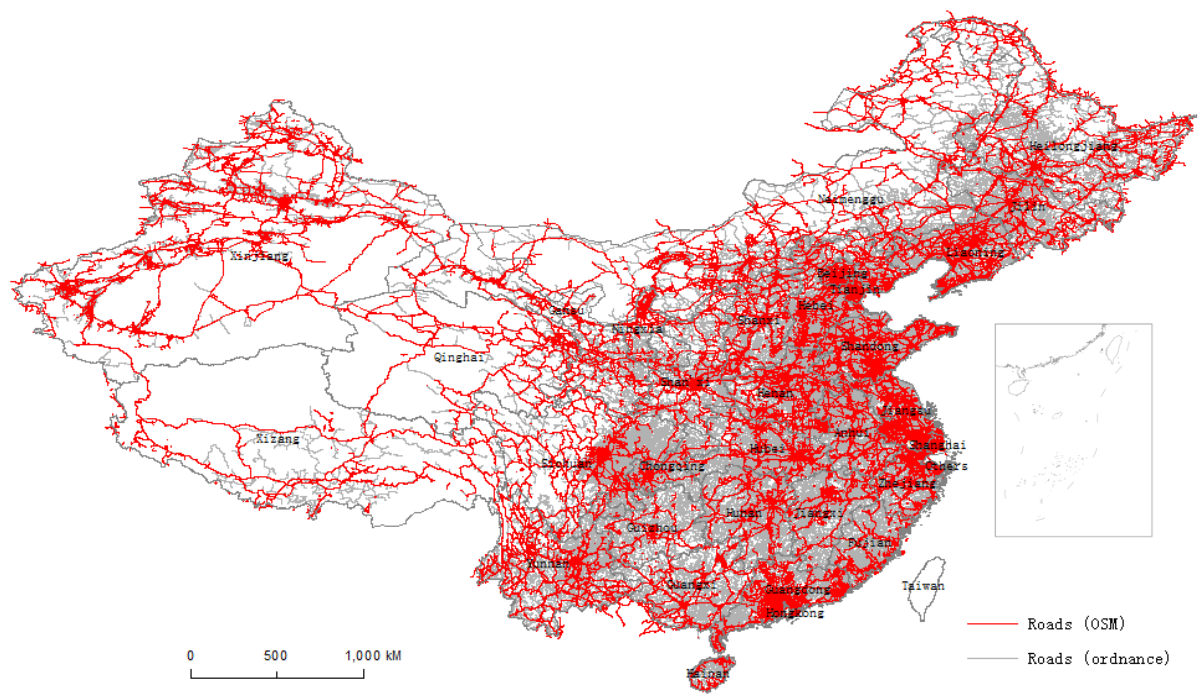

*POIs*

A total of 5,281,382 geo-tagged POIs are synthesized from a leading online business catalogue in China: Sina Weibo. The initial twenty POI types are aggregated into eight more general categories (Table 1): Commercial sites account for most POIs, followed by office building/space, transportation facilities, and government buildings. POIs labeled as 'others' are used in estimating land use density, but removed from land use mix analyses as POIs of this type are not well classified and would conduce to the estimation of the degree of mixed land use. We manually check randomly sampled POI data points to gauge and ensure the overall data quality. Our empirical framework is extensible in the sense that POI counts can be replaced by other human activity measurements, ranging from the more conventional land use cover derived from remote sensing images to online check-in service data (e.g., Foursquares).

---

[4] Roads in the ordnance map were partially covered by roads in OSM. According to our careful check, almost all roads in OSM were meanwhile in roads in the ordnance map.

Table 1 POIs types and aggregated information

| Type | Abbreviation | Counts |
| --- | --- | --- |
| Commercial sites | COM | 2,573,862 |
| Office building/space | OBS | 677,056 |
| Transport facilities | TRA | 561,236 |
| Others | OTH | 534,357 |
| Government | GOV | 468,794 |
| Education | EDU | 285,438 |
| Residence communities | RES | 167,598 |
| Green space | GRE | 13,041 |

*Other data*

DMSP/OLS (1-km spatial resolution; Yang et al. 2013) and GLOBCOVER (300-m spatial resolution; Bontemps 2009) remote sensing images are obtained for model validation, as we will compare parcels identified by our empirical framework with those generated from remote sensing images. In addition, for benchmarking purpose, manually generated parcel data for Beijing is gathered from Beijing Institute of City Planning (BICP). In our analyses, we bear in mind that the spatial resolutions of parcels generated from different methods may differ.

**Methods**

*Delineating parcel boundaries*

The working definition of a parcel in our framework is a continuously built-up area bounded by roads. Identifying land parcels and delineating road space are therefore *dual* problems. In other words, our approach begins with the delineation of road space, and individual parcels are formed as polygons bounded by roads.

The delineation of road space and parcels is performed as follows: (1) All OSM road data are merged as line features into a single data layer; (2) individual road segments are trimmed with a threshold of 200m to remove hanging segments; (3) individual road segments are then extended on both ends for 20m to connect nearby but topologically separated lines; (4) road space is generated as buffer zones around road networks. A varying threshold ranging between 2-30 m is adopted for different road types, considering both surface conditions and different levels of roads (e.g., national highways and local streets); (5) parcels are delineated as the space left when road space is removed; and (6) a final step involving overlaying parcel polygons with administrative boundaries to determine which individual parcels belong to which cities. Parameters used in these steps are determined pragmatically and based on practitioners' experiences, while bearing in mind the topological errors of OSM data.

*Calculating density for all parcels*

We define land use density as the ratio between the counts of POIs in/close to a parcel to the parcel area[5]. We further standardized the density to range from 0 to 1 for better inter-city and intra-city density comparison using Equation (1):

$$d = \frac{\log d_{raw}}{\log d_{max}} \quad (1)$$

where $d$ is standardized density, $d_{raw}$ and $d_{max}$ correspond to density of individual parcels and the nation-wide maximum density value[6]. As mentioned previously, other measures (e.g., online check-ins and floor area ratio) can substitute POIs and approximate the intensity of human activities.

*Selecting urban parcels*

The next step selects urban parcels from all generated parcels. The total urban land of individual cities was gathered from MOHURD (2013). We employ a vector-based constrained cellular automata (CA) model to identify urban parcels in individual cities[7] (Zhang and Long 2013). More specifically, we use the CA model to predict individual parcels' possibility of being urban, and the total urban land is used as constraints for the aggregated amount of urban parcels.

---
[5] POIs within the buffered road space were accounted by their closest parcels in our experiment.
[6] The unit is the POI count per $km^2$. For parcels with no POIs, we assume a minimum density of 1 POI per $km^2$.

In the CA model, each parcel is regarded as a cell in CA, and the cell status was 0 (urban) or 1 (non-urban). The CA model essentially simulates the urban development. On the onset of the simulation, all cells are set to be rural. During each step during the simulation, whether a parcel is converted to 'urban', i.e., the probability of being urban, depends on two factors (Li and Yeh 2002): Firstly, the proportion of neighboring parcels that are urban. In our empirical operationalization, the neighborhood of a parcel includes all parcels within a 500m radius; and secondly, individual parcels' attributes such as size, compactness, and the POIs density. These three attributes are combined using a logit function to influence individual parcels' probability of being urban (Wu, 2002). We then multiply the two factors (neighborhoods and parcel attributes) to determine whether the final probability is over a predefined threshold. In other words, a parcel surrounded by many urban parcels and with certain attributes would have more chance to be identified as an urban parcel in the simulation. Figure 4 provides a visual illustration of our CA model, where the final probability for being selected as an urban parcel for parcels A, B and C is 0.6 (0.8*6/8), 0.3 (0.6*4/8), and 0.225 (0.9*2/8) respectively. With a threshold of 0.5, the only parcel that would be selected as urban in our simulation is parcel A. The algorithm stops when the total area of urban parcels reaches total urban land.

---

[7] Each city has its own constrained CA model for identifying urban parcels.

Figure 4 Examples of identifying urban parcels using constrained CA. The parcels under investigations are labelled as A, B, and C, and the rectangles in black refer to individual parcels' neighborhoods. The numbers in the brackets denote individual parcels' probability of being urban based on their attributes.

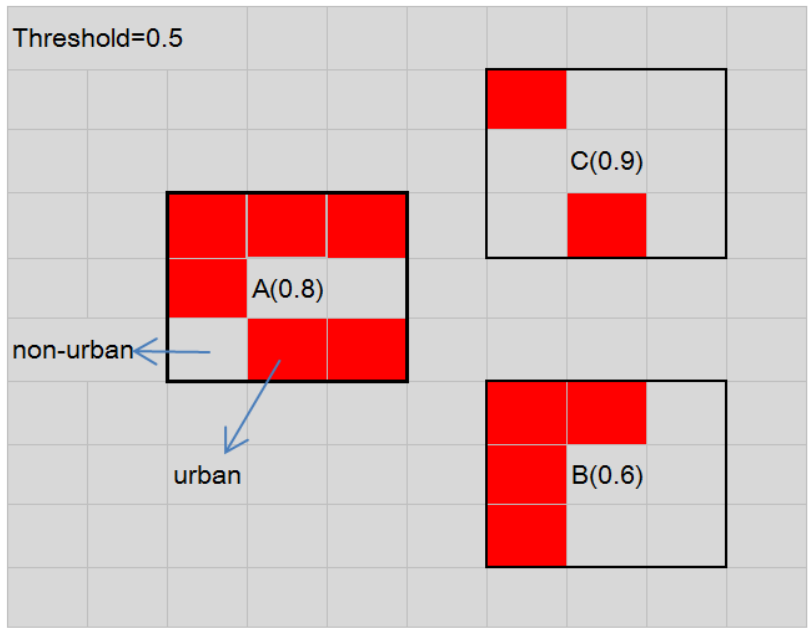

In order to determine the parameters in the aforementioned logit function, we perform a logistic regression based on parcels manually prepared for the city of Beijing (12,183 km$^2$; Yanqing and Miyun counties in the Beijing Metropolitan Area are not included). Each parcel is treated as a sample in the logistic regression: there are 125,401 samples in total, 57,817 of which are urban. The observed status of being urban/non-urban is regressed on individual parcels' size, compactness, and POI density. The overall precision of logistic regression is 74.2%, and all parameters are statistically significance. The derived parameters, which capture the relatively importance of different attributes, are incorporated into the constrained CA models for all cities in China[8]. Our constrained CA model is then validated with data for Beijing. The overall accuracy of 78.6% in terms of parcel counts implies the applicability of our CA model in identifying urban parcels from all parcels generated in a city.

*Inferring dominant urban function and land use mix for selected urban parcels*

Urban function for individual parcels is identified by examining dominant POI types within the parcels. A dominant POI type within a parcel is defined as the POI type that has accounted for more than 50% of all POIs within the parcel. For example, if 31 out of 60 POIs within a parcel are labeled as 'office building/space', the urban function for that parcel will be assigned as 'office'. Note that not all parcels would have a dominant urban function.

---

[8] We admit the heterogeneity of weights in various cities, however we do not have existing parcels for other cities at the time of this research.

As a supplement measurement for the dominant function, we computed a mix index to denote the degree of mixed land use (Frank et al. 2004). The mixed index (*M*) of a land parcel is calculated using Equation (2):

$$M = -\sum_{i=1}^{n}(p_i \times \ln p_i) \qquad (2)$$

where *n* denotes the number of POI types, and $p_i$ is the proportion of POI type *i* among all POIs in the parcel. This index has been used previously to understand evolving travel mode choice and public health outcomes, as well to study changing senses of community (Manaugh and Kreider 2013).

*Validations*

Our parcel identification and characterization results are validated at two spatial levels: At a first more fine spatial scale (i.e., parcel level), we compare the geometry and attributes of urban parcels generated by our program with those identified manually in the conventional approach. Due to data availability, this fine scale comparison is only performed for the city of Beijing (i.e., the aforementioned BICP data). Since urban parcels for Beijing was collected in 2010 with a total urban area of 1677.5 km$^2$, we re-ran the constrained CA model in Beijing using this total number and regenerated urban parcels for Beijing[9].

In order to (partially) remedy the limited availability of manually collected parcel data, we perform another set of validations at the aggregated level (i.e., regional level). In a first analysis, we compare the overall distribution of urban parcels identified from OSM and Ordnance Survey data. To ensure the comparability of urban parcels from both approaches, we use road networks from ORDNANCE in place of OSM roads and re-run our program to identify urban parcels. As Ordnance Survey data report the actual roads, thus according to our working definition of parcels, parcels generated with ORDNANCE data should correspond more closely to real-world parcels. In other words, parcels generated based on ORDNANCE data are used to benchmark the validity of OSM-based product. While this first aggregated analysis focuses on OSM's data quality, a second analysis at the aggregated level evaluates the algorithm itself. This second aggregated analysis is conducted by comparing ORDNANCE-based urban parcels with 'urban patches' in the remote sensing products (GLOCOVER and DMSP/OLS).

---

[9] The urban area of the city of Beijing was 1445.0 km$^2$ in 2012 (MOHURD, 2013), which was less than that of urban parcels prepared by BICP (1677.5 km$^2$). Such inconsistence between official yearbooks (i.e., MOHURD reports) and geospatial data (i.e., BICP data) in China is not rare.

## Results

*Parcel characteristics*

We run the proposed constrained CA model in all 654 cities. Our method generates exceedingly large parcels (i.e., individual parcels that would exceed the total urban area constraints) for cities with limited OSM data. We adopt a pragmatic threshold of ten parcels and deem the 297 cities with ten or more urban parcels as 'successfully' processed by our algorithm (Figure 5). Due to the city's sheer size – roughly the same size as Austria– Chongqing was the only MD-level city absent from this group of successfully processed cities. All SPC cities, as well as half of the medium-to-small cities at the PLC and CLC levels have result in more than ten urban parcels.

Figure 5 All generated parcels and urban parcels in China (a, spatial distribution; b, the profile of "successfully processed" cities)

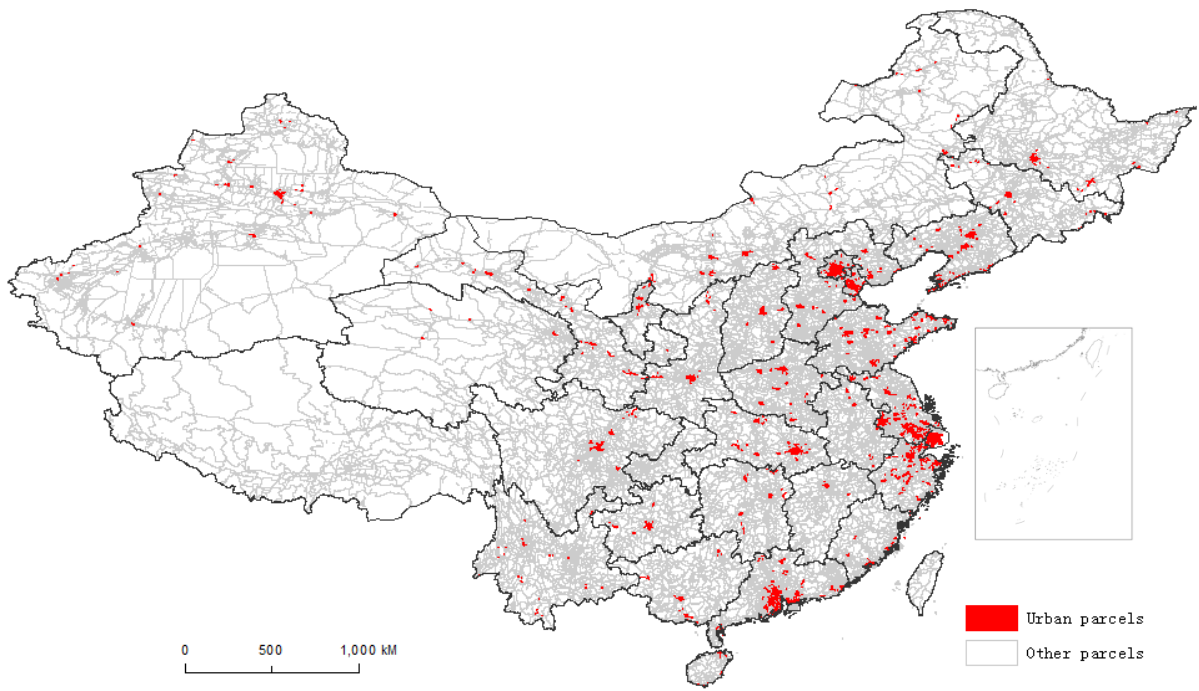

(a)

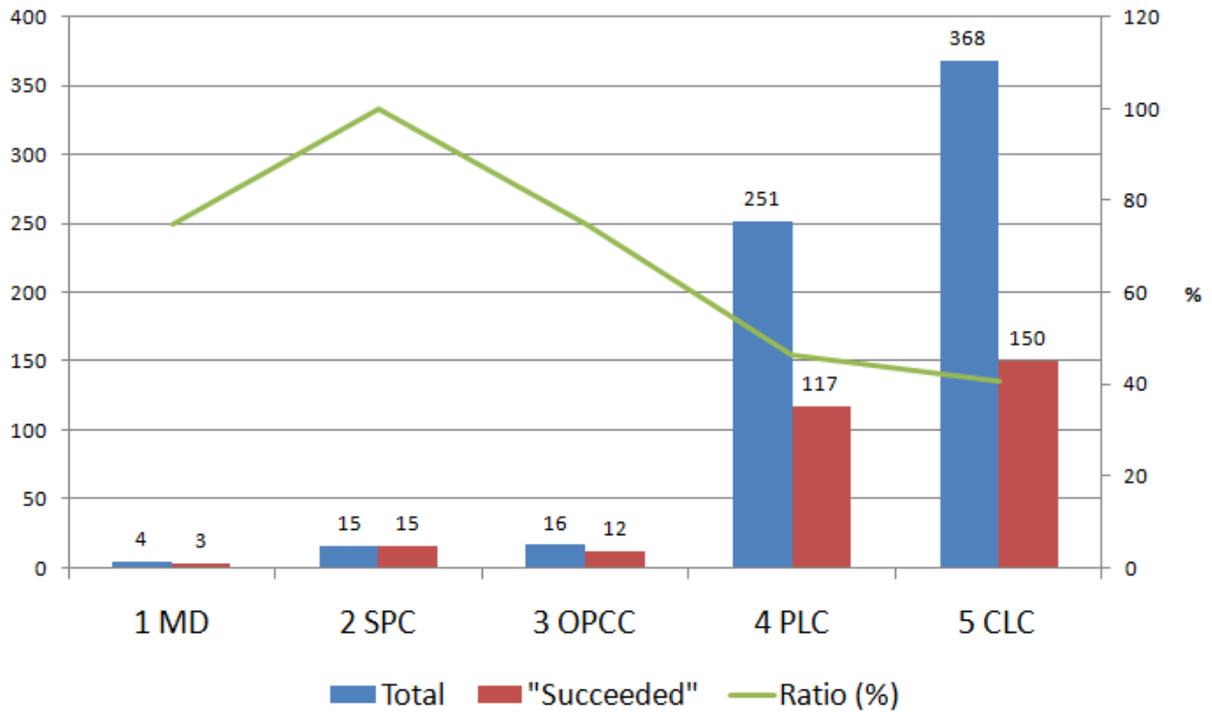

(b)

A total of 232,145 parcels are identified for these 297 cities (Figure 5), and 82,645 out of all generated parcels are labeled as 'urban' (total urban area 25,905 km$^2$). The average number of urban parcels for MD, SPC, OPCC, PLC and CLC cities are 1411, 407, 199, 79 and 26, respectively. As discussed previously, cities with more population and higher administrative ranks (e.g., Beijing, the national capital; Nanjing, a provincial capital; and Qingdao, a sub-provincial level city) tend to have more detailed OSM road network and subsequently greater number of parcels.

For all urban parcels, we calculate (1) land use density; (2) urban function; and (3) land use mix degree. Figure 6 illustrates the results for five representative cities. Density among parcels within a city or across cities could be compared in terms of inferred and standardized density attributes. Urban function and land use mix measurements point to substantive mixing of land use. More specifically, 58,915 (71.3%) out of the 82,645 urban parcels have 'dominant' urban functions (Figure 6), including 12,448 residential parcels, 11,353 commercial parcels, 9,797 Office building/space parcels, and 3,301 government parcels. Moreover, the average land mix degree for all urban parcels in 297 cities is approximately 0.66 (with a maximum of 1). The generated parcel data are distributed freely online at www.beijingcitylab.com.

Figure 6 The generated parcels and their attributes in typical cities of China

| | Beijing | Nanjing | Changsha | Weifang | Gongzhuling |
|---|---|---|---|---|---|
| **City level** | 1 MD | 2 SPC | 3 OPCC | 4 PLC | 5 CLC |
| **Density** 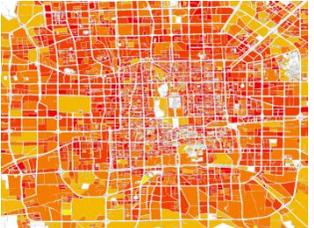 | 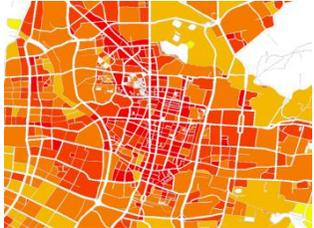 | 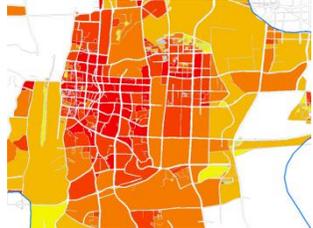 | 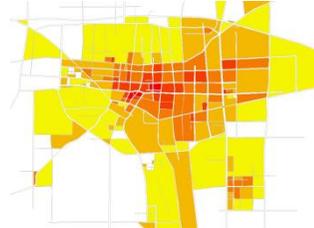 | 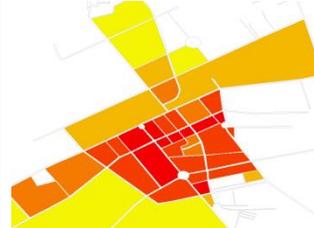 | 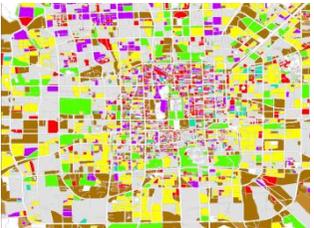 |
| **Function** 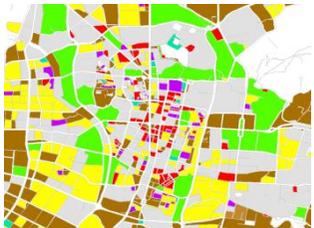 | 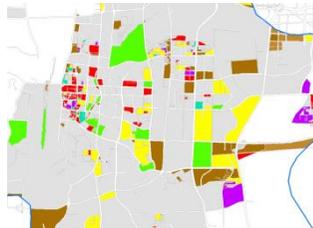 | 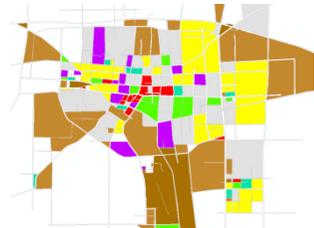 | 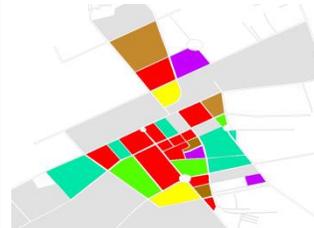 | | |

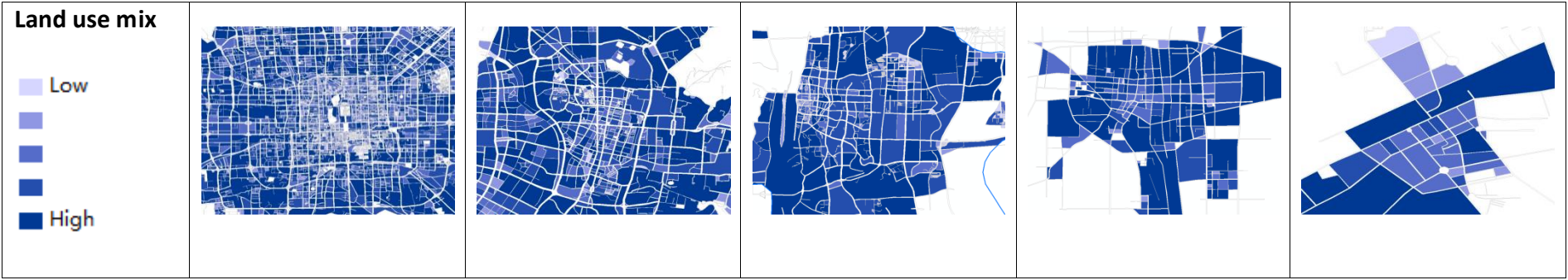

*Parcel validation*

For validation at the parcel level, Table 2 summarizes the comparison of parcels generated by our approach and those contained in the BICP Beijing parcel data. Table 2 suggests that OSM-based approach generally produce larger parcels, due to the lack of information about tertiary and more detailed roads in the OSM dataset[10]. Nevertheless, the overlapped area between urban parcels in the OSM and BICP data accounts for 71.2% of all OSM-based urban parcels, suggesting that both datasets largely capture the same geographic distribution of land use activities. In addition, we decompose the city of Beijing into sub-regions bounded by major ring roads, and calculate the proportion of parcels falling into individual sub-regions. The proportion of parcels falling into sub-regions between ring roads is consistent across both datasets. We also compare the size distribution of parcels, both of which are showing lognormal distribution with similar mean values.

Table 2 Comparison of selected OSM and BICP parcels in Beijing (R=ring road)

| Parcels | Parcel count | Average size (ha) | Overlapped with BICP | Spatial distribution (in terms of area, $km^2$) | | | | | |
|---|---|---|---|---|---|---|---|---|---|
| | | | | Within R2 | R2-R3 | R3-R4 | R4-R5 | R5-R6 | Beyond R6 |
| OSM | 7,130 | 17.2 | 1194.2 km2 (71.2%) | 42.5 | 74.0 | 113.4 | 263.5 | 666.5 | 519.9 |
| BICP | 57,818 | 2.9 | - | 48.6 | 69.7 | 99.8 | 229.5 | 687.9 | 544.4 |
| OSM/BICP | 0.12 | 5.93 | - | 0.87 | 1.06 | 1.14 | 1.15 | 0.97 | 0.95 |

Furthermore, density and urban functions of OSM-based urban parcels in Beijing are compared with other data sources. With the same OSM-generated parcel boundaries, we calculate development density for individual parcels (a total of 7,130 parcels[11]) based on (1) building information such as floor area gathered from BICP for 2008; and (2) POI data, as building information is the common data for inferring development density. The Pearson correlation coefficient between development densities calculated in two different ways is 0.858, suggesting that POI data could be used as a proxy for urban density. As POI types and land use types in BICP data were not totally aligned with each other, we limited our comparison to OSM parcels with a dominating residence function and residential parcels in BICP. We overlaid residential parcels in OSM and BICP, and the overlapping area is 211.5 $km^2$ (56.3% out of total 375.6$km^2$ OSM-based residential parcels). In other words, the parcel level validation suggests that, despite only using data from online sources, our OSM-based approach often produce reasonably good approximations of data produced by the conventional manual method.

---

[10] Parcels by ORDNANCE in Beijing were similar with those by planners in BICP in terms of parcel size.
[11] Density for BICP parcels is calculated based on floor space rather than POI. Floor space information was limited to the parcels within the six ring road of Beijing, which was used in comparison.

Validation at the aggregated regional level is performed by comparing urban parcels generated by OSM and the Ordnance Survey data ('ORDNANCE') in 297 cities where substantive numbers of urban parcels are identified (Table 3). Urban parcels based on ORDNANCE are generated and selected using the same parcel generation and selection methods as applied to the OSM data. Table 3 suggests that OSM-based approach tends to generate parcels of larger size, again due to the relative sparseness of road networks in OSM data. The match degree between urban land by OSM and ORDNANCE is 58.1%, calculated as the ratio of the area of overlapping urban parcels to the area of all OSM-based urban parcels. When we disaggregate the overlapping results in each city level, the ratio for MC, SPC and OPCC is around 70% and the ratio for FLC and CLC is around 45%. This, following the comparison on road networks in both datasets in Figure 2, further confirms the data completeness of OSM in big cities was much better than that in small cities in China.

Table 3 The comparison of OSM and ORDNANCE urban parcels for 297 cities

| Data | Urban area (km$^2$) | Parcel count | Average parcel/patch size (ha) | Intersected with ORDNANCE (km$^2$) |
|---|---|---|---|---|
| OSM | 25,905 | 82,645 | 31.3 | 15,053 |
| ORDNANCE | 25,670 | 260,098 | 10.0 | - |

Additionally, as the errors in OSM-generated parcels may come from (1) errors in the raw OSM data; and (2) errors in our empirical framework, we attempt to single out pitfalls in our empirical framework. In this regard, we cross-reference ORDNANCE-based parcels with remote-sensing based parcels. On the one hand, we expected ORDNANCE-based parcels would be less plagued by data issues that linger over the applicability of OSM, and thus reflect the effectiveness of our algorithms. On the hand, we deem urban areas identified from remote sensing images as the 'ground truth'.

More specifically, we compare the urban parcels based on ORDNANCE with the 300m-resolution urban area of China in GLOBCOVER (Bontemps, 2009), as well as 1 km-resolution urban area of China from DMSP/OLS in 2008 (Yang et al. 2013). We quantify the overlapping between urban parcels generated based on ORDNANCE data and 'urban patches' identified in the remote sensing products.

There were 21,553 km$^2$ urban lands in ORDNANCE (54.2%) overlapping with those of DMSP/OLS. This overlapping percentage would rise to 60%, if we assume road spaces that are not accounted for in ORDNANCE were covered by DMSP/OLS. We also found the comparison results between ORDNANCE and GLOBCOVER were robust, though less promising than those between ORDNANCE and DMSP/OLS. The overlapped area of the two datasets was 19,501 km$^2$, 49.5% of urban area in GLOBCOVER and 43.6% of urban area in DMSP/OLS. Considering the sheer size of our study area and limited data sources, we believe that these overlapping percentages are reasonably good.

We note that these overlapping percentages are achieved even though remote sensing images only capture the ground truth at a rather coarse-scale. For example, Table 4 suggests that an average urban parcel was around 300m-by-400 m, much smaller than the average size of an 'urban patch' in GLOBCOVER or DMSP/OLS. In addition, the time lag between ORDNANCE and DMSP/OLS, to certain extent, underestimated the overlapping ratio. Still, overlapping percentage might be hampered by the inconsistency between spatial resolutions of the two datasets. In the meantime, we are striving to collect fine scale urban land use data for the entire China, which would enable more detailed validation of our method.

Table 4 The comparison of urban parcels/patches in various data for 627 cities

| Data | Year | Spatial resolution | Urban area (km$^2$) | Parcel/patch count | Average parcel/patch size (ha) | Intersected with ORDNANCE (km$^2$) |
|---|---|---|---|---|---|---|
| ORDNANCE | 2011 | - | 39746 | 350102 | 13.0 | - |
| DMSP/OLS | 2008 | 300 m | 44720 | 1293 | 3458.6 | 21553 |
| GLOBCOVER | 2009 | 1 km | 39389 | 12515 | 314.7 | 15206 |

## Conclusions

Aiming at the paucity of parcel data in cities of the developing world, our study proposes a novel and extensible empirical framework for the automatic identification and characterization of parcels using OSM and POIs data. Our analysis represents a preliminary attempt to use volunteered GIS data to identify and characterize urban parcels in China. Empirical results suggest that OSM and POIs could help to produce reasonably good approximation of parcels identified from conventional methods, thus making our approach a useful supplement. The bottom line, however, is that we may argue for more ways of identifying parcels, not just newer, faster, or more efficient ways. In fact, as we have enumerated in the review section, parcels have been conventionally produced in a number of different ways. Our new web-based method thus provides more opportunities and alternatives to characterize parcels and generate insights.

More specifically, the contribution of this paper lies in the following aspects: Firstly, we propose a robust and straightforward approach to delineating parcels, identifying urban parcels, and characterizing parcel features, using ubiquitously available OSM data. Secondly, we employ a novel approach that incorporates a vector-based cellular automata model into the identification of urban parcels. Thirdly, our approach has been applied to hundreds of cities in China, and could possibly be extended to generate parcel data for other areas that lack of conventional data sources.

The final product of our project is a dataset containing fine scale urban parcels with detailed features for 297 Chinese cities. This dataset can be applied to but not limited to the following aspects: Firstly, the dataset can be updated periodically and provides parcel maps for urban planning and studies in places where digital infrastructure development is weak. For example, official parcel data for Beijing are generally updated every three years and our approach would enable updating on a yearly basis to capture rapid growth in Chinese cities. Secondly, the dataset can serve as the base for emerging vector-based urban modeling, e.g., vector-based cellular automata models and agent based models (Stevens and Dragicevic 2007; Jjumba and Dragicevic, 2012). Urban parcels generated by our approach would enable us to establish large-scale urban expansion models for large geographic areas (e.g., an entire nation) at parcel level. Such urban expansion models would open up new avenues for fine-scale regional growth management but were technically impossible without parcel data. Our attempt to establish such parcel-level national-scale urban expansion model will be reported in a related paper. Thirdly, parcel attributes such as urban functions and land use intensity provide useful measurements for urban analysts to examine *inter alia* quality of life, urban growth, and land use changes (Frank et al. 2010). As a sign of the usefulness of our project, though our dataset has been released for a very brief period of time, many planning projected and urban analyses have reportedly explore and use our parcel data: our parcel dataset has been downloaded more than 1500 times and we received over 100 comments in its first week of release. In the past, planning professionals in China have less access to land use data at such fine spatial scale. Fourthly, the generated parcels could be used as spatial units for consolidating other spatially referenced data, e.g., geotagged photos, transportation smart card records, taxi trajectories and mobile phone traces. The estimation of urban function, density and land use mix would be improved by integrating different data sources.

As discussed previously, we aware that our method is likely to be susceptible to the caveats of crowd-sourced data (Sui et al. 2013). For example, we note that, in addition to data completeness, other issues such as data accuracy, data 'vandalism', temporal inconsistency between data uploaded at different times, and flexible taxonomy in meta data may also affect OSM data quality. All these caveats need to be handled with caution when open data such as OSM data are applied (Haklay et al. 2010; Neis et al. 2012). Because the general limitations and setbacks of using open and crowd-sourced data to study urban dynamics have been detailed elsewhere (Elwood et al. 2012; Liu et al. 2013; Sui et al. 2013; Sun et al. 2013), we conclude by noting limitations and possible future research avenues that are specific to our empirical framework. A first limitation of our approach is that OSM road networks are relatively sparse in many cities (especially those at the lower end of the administrative hierarchy) and lead to unrealistic large urban parcels. This deficiency is likely to be alleviated by the ever-increasing coverage and quality of OSM data in China (Figure 1). Techniques for parcel subdivision would be an alternative solution for generating more realistic urban parcels in small cities in China (Aliaga et al. 2008). A second limitation is related to the use of POIs for estimating land use density. Our current approach focuses on the quantity rather than quality of individual POIs (e.g., a large department store and a small convenience store are treated equally). Possible improvements include the incorporation of online check-in data (e.g., Jiepang and SinaWeibo – Chinese equivalents of Twitter and Foursquares, respectively), taxi trajectories, and transportation smart card records to supplement inferring land use intensity. Thirdly, the current fine-scale validation is limited to the city of Beijing, and our method needs to be validated and refined with real-world parcel data in more cities. As already mentioned, more efforts need to be put into data and model validation. Lastly, the CA model can be improved by incorporating more constraints like accessibilities to main roads and city centers, as well as exclusive development zones.

\*\*\* \*\*\* \*\*\*

As the generated parcel data are freely available online[12], the online data distribution and visualization offer a new avenue to validate and improve our methods: crowd-sourced validations (Fritz et al. 2012). More specifically, individual users, with their local knowledge and experiences, would identify and report geometric and/or thematic errors of parcels, which in turn would be incorporated and used to fine-tune our CA model.

---

[12] Beijing City Lab, Data15, http://www.beijingcitylab.com/data-released-1/data1-20/.

**Acknowledgments:** Authors contribute equally to this article. The first author would like to acknowledge the financial support of the National Natural Science Foundation of China (No.51408039), and the second author thanks the support of a seed grant from The University of Hong Kong. Any errors and inadequacies of the paper remain solely the responsibility of the authors.

## References

Alberti M, Booth D, Hill K, Coburn B, Avolio C, Coe S, Spirandelli D,2007, "The impact of urban patterns on aquatic ecosystems: an empirical analysis in Puget lowland sub-basins" Landscape and urban planning **80(4)** 345-361

Aliaga D G, Vanegas C A, Beneš B 2008, "Interactive example-based urban layout synthesis", In *ACM Transactions on Graphics (TOG)* ACM **27 (5)** 160

Barnsley M J, Barr S L, 1996, "Inferring urban land use from satellite sensor images using kernel-based spatial reclassification" *Photogrammetric Engineering and Remote Sensing* **62(8)** 949-958

Beijing Institute of City Planning, 2010 *Existing land use map of Beijing*. Internal Working Report

Beresford A R, Stajano F, 2003, "Location privacy in pervasive computing" *Pervasive Computing IEEE* **2(1)** 46-55

Bontemps S, Defourny P, Van Bogaert E, Arino O, Kalogirou V, Perez J R,2009 *GLOBCOVER: Products Description and Validation Report*

Cheng J, Turkstra J, Peng M, Du N, Ho P, 2006, "Urban land administration and planning in China: Opportunities and constraints of spatial data models" *Land Use Policy* **23(4)** 604-616

Elwood S, Goodchild M, Sui DZ, 2012, "Researching Volunteered Geographic Information: Spatial data, geographic research, and new social practice" *Annals of the Association of American Geographers* **102(3)** 571-590

Erickson A, Rogers L, Hurvit P, Harris J, 2013, "Challenges and Solutions for a Regional Land Use Change Analysis" *Proceedings of ESRI*, Available online at http://proceedings.esri.com/library/userconf/proc06/papers/papers/pap_1472.pdf (accessed on 26 November 2013)

Feng Y, Liu Y, Tong X, Liu M, Deng S, 2011, "Modeling dynamic urban growth using cellular automata and particle swarm optimization rules" *Landscape and Urban Planning* **102(3)** 188-196

Frank L D, Andresen M A, Schmid T L 2004, " Obesity relationships with community design, physical activity, and time spent in cars" *American Journal of Preventive Medicine* **27(2)** 87-96

Frank L D, Sallis J F, Conway T L, Chapman J E, SaelensB E, Bachman W 2006, "Many pathways from land use to health: associations between neighborhood walkability and active transportation, body mass index, and air quality" *Journal of the American Planning Association* **72(1)** 75-87

Frank L D, Sallis J F, Saelens B E, Leary L, Cain K, Conway T L, Hess P M, 2010, "The development of a walkability index: application to the Neighborhood Quality of Life Study" *British journal of sports medicine* **44(13)** 924-933

Fritz S, McCallum I, Schill C, Perger C, See L, Schepaschenko D, van der Verde M, Kraxner F, Obersteiner M, 2012, "Geo-Wiki: An online platform for improving global land cover" *Environmental Modelling & Software* **31** 110-123

Girres J F, Touya G, 2010, "Quality assessment of the French OpenStreetMap dataset" *Transactions in GIS* **14(4)** 435-459

Goodchild M F, 2007," Citizens as sensors: the world of volunteered geography" *GeoJournal* **69(4)** 211-221

Goodchild M F, 2008, "Spatial accuracy 2.0", In J.-X. Zhang and M.F. Goodchild, editors, Spatial Uncertainty, *Proceedings of the Eighth International Symposium on Spatial Accuracy Assessment in Natural Resources and Environmental Science*s, Volume 1. Liverpool: World Academic Union, pp. 1–7

Haklay M, 2010, "How good is volunteered geographical information? A comparative study of OpenStreetMap and Ordnance Survey datasets" *Environment and Planning B: Planning & design* **37(4)** 682

Haklay M, Weber P, 2008, "OpenStreetMap: User-generated street maps" *Pervasive Computing, IEEE* **7(4)** 12-18

Haklay M, Basiouka S, Antoniou V, Ather A, 2010, "How many volunteers does it take to map an area well? The validity of Linus' Law to Volunteered Geographic Information" *The Cartographic Journal* **47 (4)** 315-322

Herold M, Scepan J, Clarke K C, 2002, "The use of remote sensing and landscape metrics to describe structures and changes in urban land uses" *Environment and Planning A* **34(8)** 1443-1458

Jabareen Y R, 2006, "Sustainable urban forms their typologies, models, and concepts" *Journal of Planning Education and Research* **26(1)** 38-52

Jiang B, Liu X, 2012," Scaling of geographic space from the perspective of city and field blocks and using volunteered geographic information" *International Journal of Geographical Information Science* **26(2)** 215-229

Jiang B, Liu X, Jia T, 2013 "Scaling of geographic space as a universal rule for map generalization" *Annals of the Association of American Geographers* (ahead-of-print)

Jjumba A, Dragićević S, 2012, "High resolution urban land-use change modeling: Agent iCity approach" *Applied Spatial Analysis and Policy* **5(4)** 291-315

Jokar Arsanjani J, Helbich M, Bakillah M, Loos L, 2013, "The emergence and evolution of OpenStreetMap: A cellular automata approach" *International Journal of Digital Earth* 1-30 (accepted)

Jokar Arsanjani J, Helbich M, Bakillah M, Hagenauer J, Zipf A, 2013, "Toward mapping land-use patterns from volunteered geographic information" *International Journal of Geographical Information Science* 1-15 (ahead-of-print)

Kressler F P, Bauer T B, Steinnocher K T, 2001, "Object-oriented per-parcel land use classification of very high resolution images", In *Remote Sensing and Data Fusion over Urban Areas*, IEEE/ISPRS Joint Workshop 2001 pp. 164-167

Leitte A M, Schlink U, Herbarth O, Wiedensohler A, Pan X C, Hu M, … Franck U, 2012, "Associations between size-segregated particle number concentrations and respiratory mortality in Beijing, China" *International Journal of Environmental Health Research* **22(2)** 119-133

Li X, Yeh A G O, 2002, "Neural-network-based cellular automata for simulating multiple land use changes using GIS" *International Journal of Geographical Information Science* **16(4)** 323-343

Liu X, Biagioni J, Eriksson J, Wang Y, Forman G, Zhu Y, 2012, "Mining large-scale, sparse GPS traces for map inference: comparison of approaches", In *Proceedings of the 18th ACM SIGKDD International Conference on Knowledge Discovery and Data Mining* pp. 669-677

Liu Y, Sui Z, Kang C, Gao Y, 2013," Uncovering patterns of inter-urban trips and spatial interactions from check-in data" arXiv preprint arXiv:1310.0282

Liu Y, Wang F, Xiao Y, Gao S, 2012,"Urban land uses and traffic 'source-sink areas': Evidence from GPS-enabled taxi data in Shanghai" *Landscape and Urban Planning* **106(1)** 73-87

Long Y, Han H Y, Yu X, 2013, "Discovering functional zones using bus smart card data and points of interest in Beijing" Beijing City Lab, Working Paper # 11, http://www.beijingcitylab.com/working-papers/wp1-20/, accessed on 26 November 2013

Long Y, Liu X, 2014, "How mixed is Beijing, China? A visual exploration of mixed land use" *Environment and Planning A* (Forthcoming)

Ma L J, 2005,"Urban administrative restructuring, changing scale relations and local economic development in China" *Political Geography* **24(4)** 477-497

Neis P, Goet M, Zipf A, 2012, "Towards automatic vandalism detection in OpenStreetMap" *ISPRS International Journal of Geo-information* **1(3)** 315-332

Ministry of Housing and Urban-rural Development of the People's Republic of China (MOHURD), 2013 *Chinese City Construction Statistics Yearbook 2012* (Beijing: China Planning Press)

Over M, Schilling A, Neubauer S, Zipf A, 2010," Generating web-based 3D City Models from OpenStreetMap: The current situation in Germany" *Computers, Environment and Urban Systems* **34(6)** 496-507

Pacifici F, Chini M, Emery W J, 2009,"A neural network approach using multi-scale textural metrics from very high-resolution panchromatic imagery for urban land-use classification" *Remote Sensing of Environment* **113(6)** 1276-1292

Pinto N N, 2012, "A cellular automata model based on irregular cells: application to small urban areas" *Environment and Planning B: Planning & Design* **37(6)** 1095-1114

Ramm F, Topf J, Chilton S, 2010 *OpenStreetMap: using and enhancing the free map of the world* (Cambridge: UIT Cambridge)

Soto V, Frías-Martínez E, 2011 "Automated land use identification using cell-phone records", In *Proceedings of the 3rd ACM international workshop on MobiArch* ACM pp. 17-22

Stevens D, Dragicevic S, 2007, "A GIS-based irregular cellular automata model of land-use change" *Environment and Planning B: Planning & Design* **34(4)** 708-724

Sui D Z, 2008, "The wikification of GIS and its consequences: Or Angelina Jolie's new tattoo and the future of GIS" Computers, *Environment and Urban Systems* **32(1)** 1-5

Sui D Z, Goodchild M, Elwood S, 2013, " Volunteered Geographic Information, the exaflood, and the growing digital divide", in *Crowdsourcing Geographic Knowledge* Eds D Sui et al. (Springer Netherlands) pp. 1-12

Sun L, Axhausen K W, Lee D H, Huang X, 2013, "Understanding metropolitan patterns of daily encounters" *Proceedings of the National Academy of Sciences* **110(34)** 13774-13779

Toole J L, Ulm M, González M C, Bauer D, 2012, "Inferring land use from mobile phone activity", In *Proceedings of the ACM SIGKDD International Workshop on Urban Computing* ACM pp1-8

Wu F, 2002, "Calibration of stochastic cellular automata: the application to rural-urban land conversions" *International Journal of Geographical Information Science* **16(8)** 795-818

Yang Y, He C, Zhang Q, Han L, Du S, 2013,"Timely and accurate national-scale mapping of urban land in China using Defense Meteorological Satellite Program's Operational Linescan System nighttime stable light data" *Journal of Applied Remote Sensing* **7(1)** 073535

Yuan J, Zheng Y, Xie X, 2012, "Discovering regions of different functions in a city using human mobility and POIs", In *Proceedings of the 18th ACM SIGKDD International Conference on Knowledge Discovery and Data Mining* ACM pp. 186-194

Zhang L, Yang W, Wang J, Rao Q, 2013, "Large-Scale Agent-Based Transport Simulation in Shanghai, China" *Transportation Research Record: Journal of the Transportation Research Board* **2399(1)** 34-43

Zhang Y P, Long Y, 2013, "Urban Growth Simulation Using V-BUDEM: A Vector-Based Beijing Urban Development Model" The conference of Spatial Planning and Sustainable Development, Beijing

Zheng S, Zheng J, 2014, "Assessing the completeness and positional accuracy of OpenStreetMap in China", in Thematic Cartography for the Society, *Lecture Notes in Geoinformation and Cartography* Eds T. Bandrova et al. (Springer International Publishing, Switzerland) pp. 171–189